# Rooting behavior of pomegranate (*Punica granatum* L.) hardwood cuttings in relation to genotype and irrigation frequency

KOCHER OMER SALIH[1], ARAM AKRAM MOHAMMED[1*], JAMAL MAHMOOD FARAJ[1], ANWAR MOHAMMED RAOUF[1], NAWROZ ABDUL-RAZZAK TAHIR[1]

[1] Horticulture Department, College of Agricultural Engineering Sciences, University of Sulaimani, Kurdistan Region, Iraq

*Correspondence details: aram.hamarashe@univsul.edu.iq

**Abstract:** The study was conducted to determine the best irrigation frequency for rooting hardwood cuttings of some pomegranate genotypes that are cultivated in Halabja province, Kurdistan Region, Iraq. The hardwood cuttings were collected from 11 genotypes, which were 'Salakhani Trsh' (G1), 'Salakhani Mekhosh' (G2), 'Amriki' (G3), 'Twekl Sury Trsh' (G4), 'Twekl Astury Naw Spy' (G5), 'Hanara Sherina' (G6), 'Kawa Hanary Sherin' (G7), 'Kawa Hanary Trsh' (G8), 'Malesay Twekl Asture' (G9), 'Malesay Twekl Tank' (G10), and 'Sura Hanary Trsh' (G11). The genotypes were subjected to irrigation applications by 1-day, 2-day, 7-day, or 10-day frequencies. Among pomegranates, G11, G6, and G7 produced 95, 90, and 83% rooting percentages, which were significantly higher than the rest of other genotypes. The lowest rooting percentages (28, 36, 38, and 40%) were found in G1, G5, G3, and G10, respectively. The effect of irrigation frequencies on the genotypes confirmed that a 7-day frequency was the best irrigation frequency to achieve the maximum rooting percentages (93, 86, 80, 73, 53, and 40%) in G6, G9, G2, G4, G3, and G1, respectively. In contrast, the minimum rooting percentage (20%) was recorded in G3 with a 1-day frequency and in G1 with 10-day frequency. In this study, it was found that the cuttings of G11, G6, and G7 had the best ability to form roots, and irrigation with a 7-day frequency was the best for the cuttings of all the 11 pomegranate genotypes investigated.

## Introduction

Pomegranate (*Punica granatum* L.) is a member of the Punicacease family, and it is the only species cultivated for fruit production which belongs to the *Punica* L. genus (Melgarejo et al., 2012). Natural pomegranate forests are vastly distributed in the north of Iran on the Caspian Seashore, and in the Zagros Mountain plain forests in Charmahal Bakhtiayari, Fars, Kurdistan, Lorestan, Baluchistan, and the southern territories of the Alborz Mountain range. For that reason, Persia is the origin place of *P. granatum* (Zarie et al., 2021). Pomegranate is a deciduous shrub that grows up to 3-4 meters and sometimes reaches 9 meters (Smith, 2014). It can withstand arid, semi-arid, and salinity conditions. Hence, pomegranate is advisable for cultivation in areas where other fruits cannot be grown because of climatic change (Singh et al., 2011). Furthermore, fruits of

*P. granatum* are an important source of antioxidants (poly-phenols: ellagic acid and punicalagin), minerals, vitamins, and tannins which collectively improve health status and immune system against many diseases, more notably the heart and cancer diseases (Sarrou et al., 2014). In addition to health benefits, pomegranate peel and seed are also used for the extraction of oil, pectin, and phenolic compounds, along with the production of biochar, biogas and bio-oil (El-Shamy and Farag, 2021). There are many cultivars of pomegranate worldwide which are frequently named depending on the place of cultivation or color of the fruit. These cultivars are exhibited in 500 genetically different cultivars and 50 of them are recognized as commercial cultivars (IPGRI, 2001). The pomegranate cultivars are generally classified according to some discernible aspects, such as taste (sweet, sweet-sour, tart, and sour), harvesting time (early, mid-season, and late), consumption pattern (juice and table fruit), and seed hardness (soft-seeded and hard-seeded) (Kahramanoglu and Usanmaz, 2016).

Pomegranate propagation is conducted through seeds for obtaining new cultivars or breeding purposes, however, this method is not supposed to propagate a certain desirable cultivar as a result of the variabilities that occurred in the seedlings grown from seeds (Prabhuling and Huchesh, 2018). For achieving true-to-type pomegranate, it is vegetatively propagated through cuttings, grafting, air layering, and micropropagation, but the more usual one is cuttings propagation (Chandra et al., 2012). Hardwood cuttings are better to obtain vigorous pomegranate trees than softwood and semi-hardwood cuttings (Saroj et al., 2008). Cultivar is among the factors that determine the rooting ability of pomegranate cuttings (Owais, 2010). In this regard, Hejazi et al. (2023) found that the rooting ability of hardwood cuttings of pomegranate varied depending on cultivars, and cuttings of pomegranate cultivars responded differently to the rooting inducer. Moreover, Aytekin Polat and Caliskan (2009) observed that rooting rate was different among pomegranate genotypes owing to the differences in parent plant physiological conditions. Apart from these, irrigation frequency is another condition that affects rooting the cuttings (Atak and Yalçın, 2015). In this context, Yeboah et al. (2011) showed variable rooting results in *Vitellaria paradoxa* Gaertn stem cuttings, and they referred that excessive irrigation negatively affected rooting performance. On the other hand, irrigation frequency is crucial for regulating water amount and energy needed for plant production in the nurseries (Hunt and McDonald, 2015). Recognizing the role of genotype and irrigation in the rooting of cuttings, the current study was carried out to estimate rooting capacity in hardwood cuttings of some pomegranate genotypes, and their response to different irrigation frequencies.

## Materials and Methods

This study was conducted at the College of Agricultural Engineering Sciences, University of Sulaimani, Kurdistan Region, Iraq. The hardwood cuttings were taken from 11 pomegranate genotypes which are cultivated in Halabja province of the Kurdistan Region of Iraq. For simplicity, the genotypes common names are coded in the Table 1 below.

*Table 1 –The pomegranate genotypes were used in the current study with the local names*

| Genotypes | Common names |
|-----------|--------------|
| G1 | Salakhani trsh |
| G2 | Salakhani mekhosh |
| G3 | Amriki |
| G4 | Twekl sury trsh |
| G5 | Twekl astury naw spy |
| G6 | Hanara sherina |
| G7 | Kawa hanary sherin |
| G8 | Kawa hanary trsh |
| G9 | Malesay twekl asture |
| G10 | Malesay twekl tank |
| G11 | Sura hanary trsh |

## Cuttings preparation

The hardwood cuttings of the pomegranate genotypes were collected from the basal part of one-year-old shoots on February 5, 2022. The cuttings were trimmed at 15±1 cm long and had about 0.6–1 cm in diameter. From each genotype, 60 cuttings were taken. After preparation, the cuttings were divided into four groups; each group contained 15 cuttings for each genotype. The 15 cuttings of each genotype in every group were divided into three replications each consisted of 5 cuttings. The cuttings were planted in the sand medium prepared in the nursery polyethylene bags (width =15 cm, height = 30 cm). Each group was arranged on a polystyrene sheet according to Randomized Complete Block Design (RCBD) in an uncontrolled greenhouse constructed on a full-sun site with 20 m length, 9 m width, and 3 m height. The greenhouse was covered with a transparent polyethylene sheet (200 μm thickness). Daily ventilation occurred by opening the doors and slots. The four groups of the cuttings were employed to four irrigation regimes at a 1-day, 2-day, 7-day, and 10-day intervals. Hand-watering was used. Throughout the experiment, the greenhouse recorded temperature ranged from 14.7°C to 28.2°C and humidity from 37% to 62%.

## Statistical analysis

After 15 weeks, the cuttings were uprooted on late May, to take the measurements. Data were collected on rooting percentage, root number, length of the longest root, length of the longest shoot, diameter of the longest shoot, leaf number, leaf area, and relative chlorophyll content. Relative chlorophyll content was taken using a CCM-200 plus chlorophyll content meter. The length of roots and shoots was taken by the metal ruler, and the diameter of shoots was measured using an electronic digital caliper. Also, the leaf area was found by the android mobile application (Petiole), and the mobile was set at 18.5 cm distance from the leaves (Kemunto et al., 2022). The distance was fixed via a colored paper clip with 1 cm². The rooting percentage calculated based on the original number of the cuttings. XLSTAT software version 2019.2.2 was used to analyze the variance and compare the means using Duncan's Multiple-Range test at (P ≤ 0.05). Pearson Correlation test, Principle Component Analysis (PCA), and Cluster Analysis using Unweighted Pair Group Method (UPGMA) were conducted by using XLSTAT software version 2019.2.2v.

## Results and discussion

### *Impact of genotypes on the hardwood cuttings under the four irrigation frequencies*

Hardwood cuttings of different genotypes of pomegranate were evaluated to determine rooting percentage and some other cutting characteristics (Table 2). The rooting percentage was found significantly different in the different genotypes. G11, G6, and G7 produced the highest rooting percentages (95, 90, and 83%), respectively, and they were different compared to other genotypes. Moderate rooting percentages (71 and 63%) were observed in G8, G2, G4, and G9, respectively. Inversely, G1, G5, G3, and G10 showed the lowest rooting percentage (28-40%), respectively. Other researchers confirmed that cuttings of the pomegranate can produce roots, but there are differences among pomegranate cultivars or genotypes to produce roots (Aytekin Polat and Caliskan, 2009; Owais, 2010; Hejazi et al., 2023). These differences may belong to the genetic or physiological variations that occur among the genotypes, which may encourage the genotypes respond differently to the prevalent condition to which the cuttings are subjected. Chater et al. (2017) found that rooting percentages in 12 pomegranate cultivars were different when subjected to the same condition.

The effect of genotypes on other characteristics of rooted cuttings of the genotypes was shown in the same table revealed that the best root number (54.62), root length (16.77 cm), and shoot length (20.13 cm) were obtained in G6, followed by G4. Moreover, the maximum leaf number (27.54), leaf area (4.32 cm$^2$), and SPAD value (12.02 SPAD) were recorded in G11, G4, and G5, respectively. In contrast, root number (29.14) and leaf number (19.83) in G5, root length (8.85 cm), shoot length (14.69 cm), and SPAD value (4.7 SPAD) in G1, and leaf area (1.81 cm2) in G7 were the minimum. Shoot diameter was not statistically different in all the studied genotypes. Adiba et al. (2022) reported that genotype had a marked role in the growth traits of pomegranate hardwood cuttings. Generally, the genotypes that resulted in the best root number and length also had the longest shoot and the largest leaf area. These are more likely due to differences in rooting time of the cuttings of the genotypes. Thus, the genotypes rooted earlier had more time to increase root number and length, in turn increasing shoot and leaf growth because a good root system leads to better water and nutrient uptake. A higher amount of available water and nutrients could be captured by a large root system as a result of occupying more soil volume (Zalesny et al., 2005). Branislav et al. (2009) observed a connection between early rooting and shoot growth of cutting genotypes of poplar. On the other hand, it might be ascribed to differences in leaf area among the genotypes, but with a low probability. Also, the genotypes with the best leaf area in the current study showed the best shoot length. The highest leaf area is vital for conducting the highest photosynthesis, which provides more photoassimilates for better growth. Leaf area is a genotype characteristic. Some genotypes genetically have large leaves and others small leaves. Khadivi and Arab (2021) indicated that leaf lengths and widths were different in pomegranates regarding genotypes.

*Table 2 - Effect of different pomegranate genotypes on hardwood cuttings traits*

| Genotype | Rooting % | Root number | Root length (cm) | Shoot length (cm) | Shoot diameter (mm) | Leaf number | Leaf area (cm) | Chlorophyll (SPAD) |
|---|---|---|---|---|---|---|---|---|
| G1 | 28 c | 33.1 cd | 6.60 c | 14.69 b | 1.65 a | 23.42 abc | 2.90 cd | 4.70 f |
| G2 | 63 b | 32.62 cd | 11.5 bc | 14.85 b | 1.55 a | 24.91 abc | 2.89 cd | 7.41 de |
| G3 | 38 c | 31.57 cd | 10.66 bc | 16.52 ab | 1.45 a | 21.47 bc | 3.23 bc | 9.98 bc |
| G4 | 63 b | 52.60 a | 13.14 b | 20.00 a | 1.63 a | 26.05 ab | 4.32 a | 9.53 bc |
| G5 | 36 c | 29.14 d | 9.42 bc | 15.53 ab | 1.64 a | 19.83 c | 2.79 cd | 12.02 a |
| G6 | 90 a | 54.62 a | 16.77 a | 20.13 a | 1.61 a | 22.64 abc | 3.94 ab | 8.57 cd |
| G7 | 83 a | 32.32 cd | 9.52 bc | 19.13 ab | 1.72 a | 25.38 abc | 1.81 e | 10.70 ab |
| G8 | 71 b | 41.78 bc | 11.89 bc | 14.75 b | 1.63 a | 23.78 abc | 3.23 bc | 8.93 c |
| G9 | 63 b | 35.89 bcd | 9.93 bc | 15.61 ab | 1.66 a | 23.74 abc | 2.41 cde | 9.60 bc |
| G10 | 40 c | 36.99 bcd | 9.54 bc | 14.95 b | 1.59 a | 24.56 abc | 2.24 de | 9.38 bc |
| G11 | 95 a | 44.74 ab | 12.69 bc | 19.27 ab | 1.62 a | 27.54 a | 2.41 cde | 6.77 e |

Means with the same letter confirm no significant differences at P ≤ 0.05 according to Duncan's multiple-range test.

*Effect of irrigation frequency on the hardwood cuttings of the genotypes*

Irrigation frequency significantly affected the rooting percentage in hardwood cuttings of some of the pomegranate genotypes (Figure 1). Increasing the irrigation frequency from 1-day to 7-day improved the rooting percentage from (26%) to (40%) in G1. Meanwhile, the 10-day reduced the rooting percentage to the minimum (20%) in the same genotype. In G2 and G6 rooting percentages were statistically similar at 1-day, 2-day, and 7-day irrigation, but 10-day caused the lowest rooting percentage in these genotypes. Furthermore, 7-day irrigation was optimal for achieving the highest rooting percentages in G3 (53%) and G9 (86%). The irrigation of 1-day resulted in a decline in rooting percentage to 20% in G3 and 46% in G9. Irrigation frequencies did not affect the rooting percentage of G5, G7, G8, G10, and G11. Irrigation frequency may influence the rooting capacity of the cutting in two possible scenarios: Firstly, frequent irrigation unfavorably promotes rooting in cuttings of some species by minimizing available $O_2$, temperature, and increasing $CO_2$ in rooting media, which leads to rotting of the cuttings (Tsipouridis and Thomidis, 2004). Secondly, frequent irrigation may leach the essential mineral elements from the cuttings and the rooting media that required for rooting in the cuttings (Owen and Maynard, 2007). In the current study, G3 and G9 probably couldn't tolerate high water in the rooting media, which may induce hypoxia (low oxygen in the media). On the other hand, low-frequent irrigation (10-day) reduced rooting percentages in G1, G2, and G6. However, in G5, G7, G8, G10, and G11, rooting was independent of irrigation frequencies. In the present study, the pomegranate genotypes differently responded to the irrigation frequencies, which would be attributed to the genetic diversity occurs among the genotypes. Generally, hardwood cuttings from the 11 genotypes of pomegranate used in this study can successfully root with a 7-day irrigation. The findings provide valuable insights for local nursery growers, helping them understand the proper interval of irrigation for the successful rooting of hardwood cuttings from 11 pomegranate genotypes. This knowledge contributes to saving water resources, labor, and energy required for operating irrigation equipment. Fulcher et al. (2016) reported that water availability in nurseries may be diminished in the coming decades by reason of urbanization, drought, water quality decreasing, and limited withdrawals owing to regulations. Consequently, the nurseries should use less water in producing plant material with the same quantity and quality by adopting efficient cultural practices and technologies.

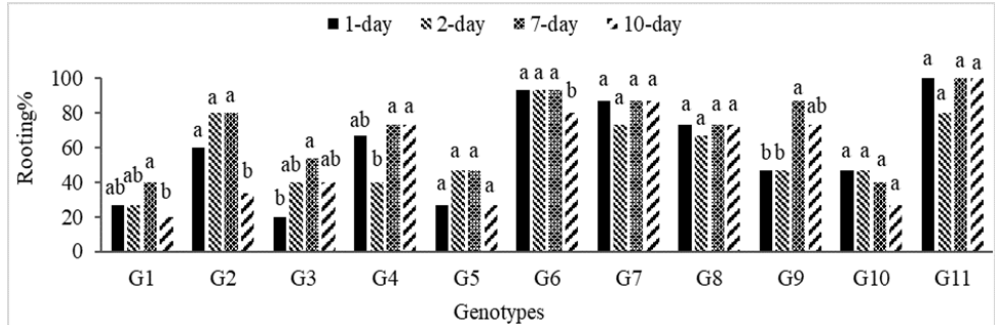

*Figure 1 - Rooting percentage of hardwood cuttings of pomegranate genotypes under 1-day, 2-day, 7-day, and 10-day irrigation frequencies. Bars with the same letter confirm no significant differences at P ≤ 0.05 according to Duncan's multiple-range test*

The data of root number (Figure 2A), root length (Figure 2B), and shoot length (Figure 2C) showed that 1-day or 2-day irrigation frequency declined the root number in G5, G6, and G7; root length in G2, G3, G5, and G7; and shoot length in G2, G5, G7, and G9. Conversely, 7-day or 10-day increased root number, root length, and shoot length in these genotypes. On the other hand, 1-day irrigation improved root number, root length, and shoot length in G10 and G11, but 7-day or 10-day minimized these parameters. The results of root traits and shoot length were likely related to differences in the response of the genotypes to the level of water in the rooting medium owing to the irrigation frequencies. It appeared that G10 and G11 were more responsive to high water content in the rooting medium, and they had the maximum root number, root length, and shoot length at the 1-day without differences in the rooting percentage. Inversely, irrigation application with 7- or 10-day resulted the best root number, root length, and shoot length in G2, G3, G5, G6, and G7. The amount of water in rooting media determines oxygen level, accumulation of ethylene, and endogenous hormone variation, which in turn alters the root number and growth of the cuttings (Owen and Maynard, 2007). In the cuttings of some species high dissolved oxygen concentration in rooting medium water proportional to better root number and growth, but in others root number and length were better when the cuttings rooted in a medium with excessive porous which hold low water amount and great oxygen level (Geneve et al., 2004). Root growth traits are interdependent with shoot growth, and this is confirmed by Pourghorban et al. (2019). Hence, in the present study, the cuttings showed more root number and length also their shoots elongated.

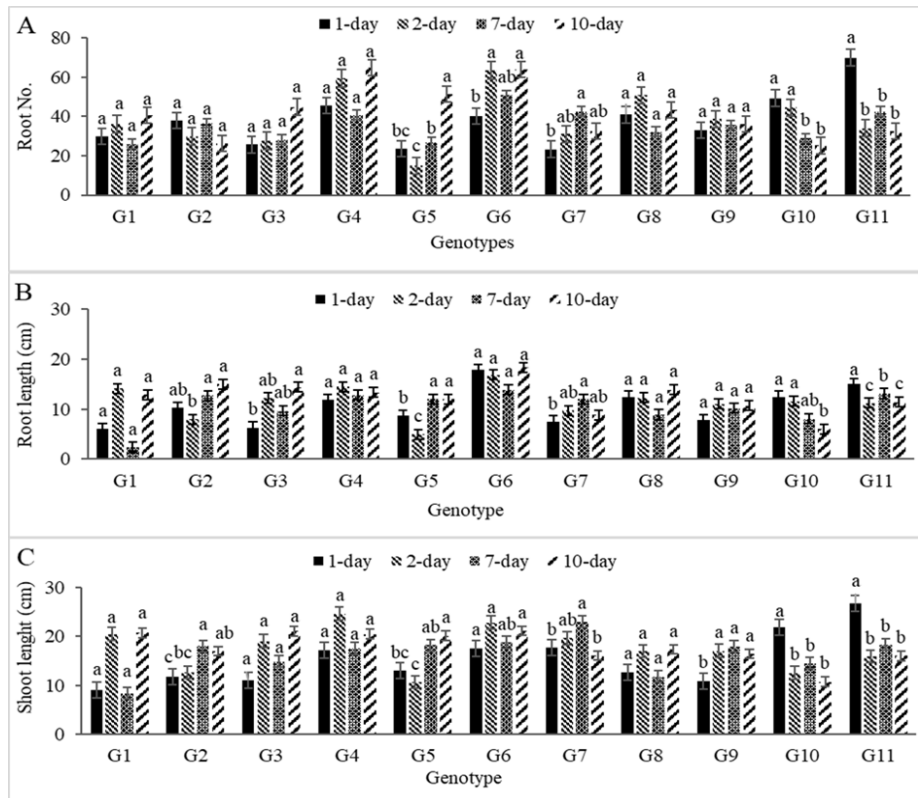

*Figure 2 - Effect of irrigation frequencies (1-day, 2-day, 7-day, and 10-day) on A) root number, B) root length (cm), and C) shoot length (cm). Bars with the same letter confirm no significant differences at P ≤ 0.05 according to Duncan's multiple-range test. The values are shown as mean ± standard error.*

Additionally, shoot diameter was not significantly different in most of the genotypes (Figure 3A). The effect of irrigation frequencies on shoot diameter was only significantly different in G7, G8, and G11. The best shoot diameters in G7 (2.01 mm) and G11 (1.88 mm) were measured at 1-day irrigation, and G8 had the best shoot diameter (2.05mm) at 2-day irrigation. Irrigation frequency at 2-day for G7 and G11, along with 7-day for G8 decreased shoot diameter. Besides, an increase in leaf number matched with expanding irrigation frequency in G2 (Figure 3B), and the highest number of leaves (29.73) were recorded in the cuttings irrigated at 10-day. At the same time, the lowest leaf numbers (21.13 and 23.22) in G2 were at 2-day and 1-day irrigation, respectively. Oppositely, leaf number in G11 declined as irrigation frequency ascended from 1-day to 10-day. 1-day was the best to rise the number of leaves (41.9) in G11, whereas 10-day was the worst (19.3). In the rest of the genotypes, the leaf number was not significantly different. Iniesta et al. (2009) discovered that irrigation and water status was directly influenced on the number of leaves. Regarding leaf area (Figure 3C), the irrigation frequencies just resulted in differences in G2 and G5. That is, 1-day and 2-day reduced leaf area in the two genotypes, but 7-day and 10-day enlarged leaf area. The best SPAD values (7, 15.5, and 17.7 SPAD) was respectively obtained in G1, G3, and G4 at 10-day irrigation (Figure 3D), while at 1-day, 2-day, and 7-day irrigation G1, G3, and G4 showed the lowest SPAD value. In G5 and G9, chlorophyll content was the highest (18.77 and 11.27 SPAD) when irrigation frequency was 2-day. However, 1-day irrigation resulted in the best SPAD value (15.9 SPAD) in G7 compared to 2-day, and 7-day, and 10-day irrigation. Moreover, SPAD value was not significantly different in G8. The differences among the genotypes in SPAD value in response to different irrigation frequencies could be attributed to the differences in genetics and physiological properties that occur among the genotypes.

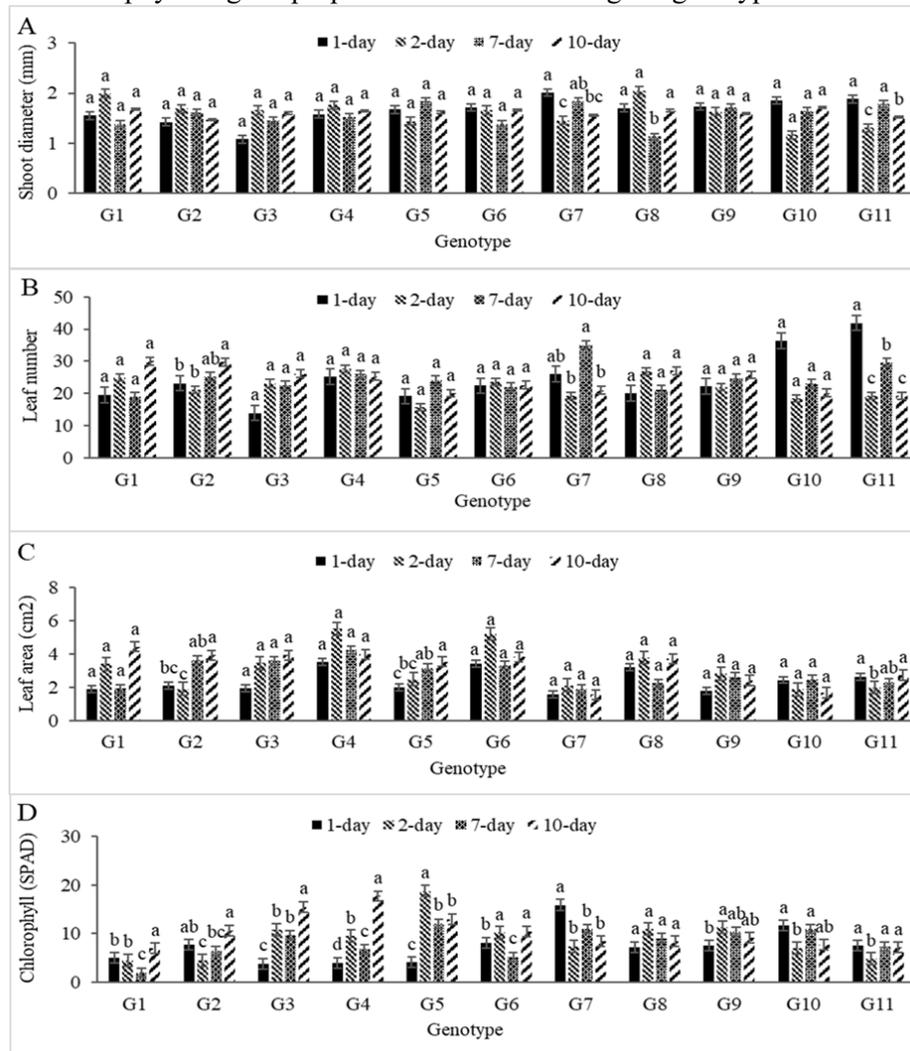

*Figure 3 – Effect of irrigation frequencies (1-day, 2-day, 7-day, and 10-day) on A) shoot diameter (mm), B) leaf number, C) leaf area (cm2), D) Chlorophyll (SPAD). Bars with the same letter confirm no significant differences at P ≤ 0.05 according to Duncan's multiple-range test. The values are shown as mean ± standard error.*

### The relationship between the genotypes and cutting characteristics

The Pearson correlation test at P = 0.05 of the means (Figure 4) indicated that a positive significant relationship existed between root number and shoot length (r=0.68, P = 0.02), and root number and leaf area (r = 0.67, P = 0.02). Additionally, root length significantly correlated with shoot length (r = 0.65, P = 0.03) and leaf area (r = 0.68, P = 0.02).

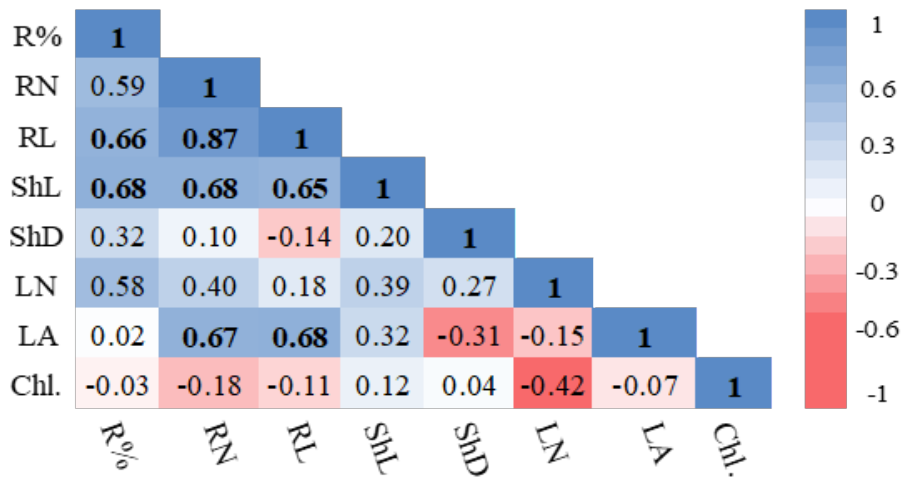

*Figure 4 – Pearson correlation test at (P ≤ 0.05) for the parameters. R%: rooting percentage, RN: root number, RL: root length, ShL: shoot length, ShD: shoot diameter, LN: leaf number, LA: leaf area, and Chl.: chlorophyll.*

Principle Component Analysis (PCA) of the variables was displayed in Figure (5) clarified that the two components PCA1 and PCA2 carried 67.73% of the variances. Component 1 (PCA1) comprised 45.03% and component 2 (PCA2) embraced 22.7%. Correspondingly, G6 and G4 are located on the positive side of the two components and strongly correlate with root number (RN), root length (RL), and leaf area (LA). G11 was a close relationship with rooting percentage (R%), leaf number (LN), shoot length (ShL), and shoot diameter (ShD). While, G1, G7, G9, and G10 are positioned on the negative side of the two components, they are in an opposite association with RN, RL, and LA. In the same manner, G2, G3, and G5 negatively related to R%, LN, ShL, and ShD. The connection between Chlorophyll (Chl.) and G11 was contradictory. RN, RL, LA, and ShL positively correlated together. Also, there was a strong correlation among R%, ShL, RN, and RL.

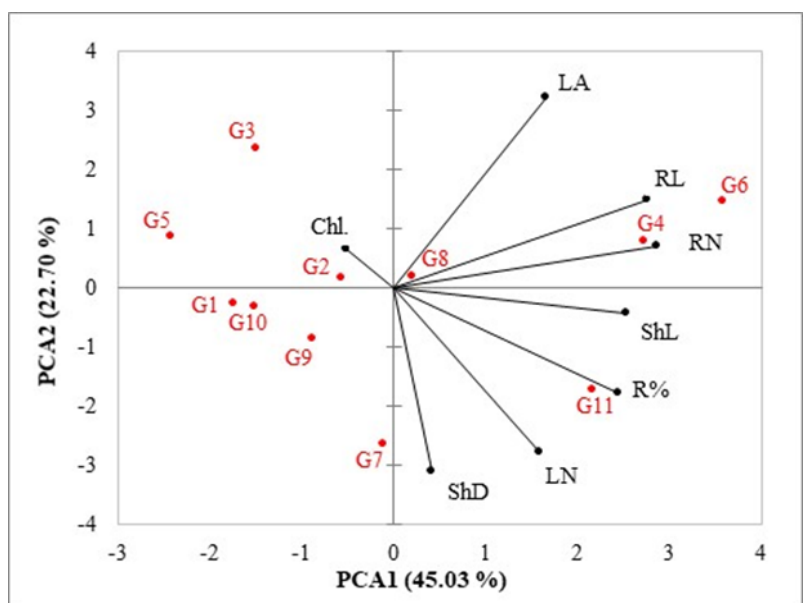

*Figure 5 – Distribution and relationship of different genotypes and cutting characteristics on PCA bi-plot. R%: rooting percentage, RN: root number, RL: root length, ShL: shoot length, ShD: shoot diameter, LN: leaf number, LA: leaf area, and Chl.: chlorophyll.*

Based on the outcomes of the cutting characteristics depending on genotype, a cluster analysis using the unweighted pair group method (UPGMA) was carried out to ascertain the similarities between the genotypes examined (Figure 6). Three major clusters were found, according to the findings. Three genotypes, divided into two distinct subclasses, made up Cluster 1 (in blue). The G1 genotype was included in the first subclass, which was distinguished by having the lowest value for root length, shoot length, and total chlorophyll content. The three genotypes with the highest total chlorophyll content were G3, G5, and G10 in the second subclass. Four genotypes (G2, G4, G8, and G9) that were divided into two subclasses made up the second cluster, which is indicated by the red color. The highest values of shoot length and total chlorophyll content were used to describe these genotypes. Three genotypes (G6, G7, and G11) presented the third cluster (in green). The highest values for shoot length and diameter set this group apart. On the basis of these results, it was determined that there was a substantial difference in cutting ability between the genotypes tested, indicating the presence of multiple genetic pools in this collection. These data are extremely important for farmers because they highlight genotypes with successful stem cutting propagation.

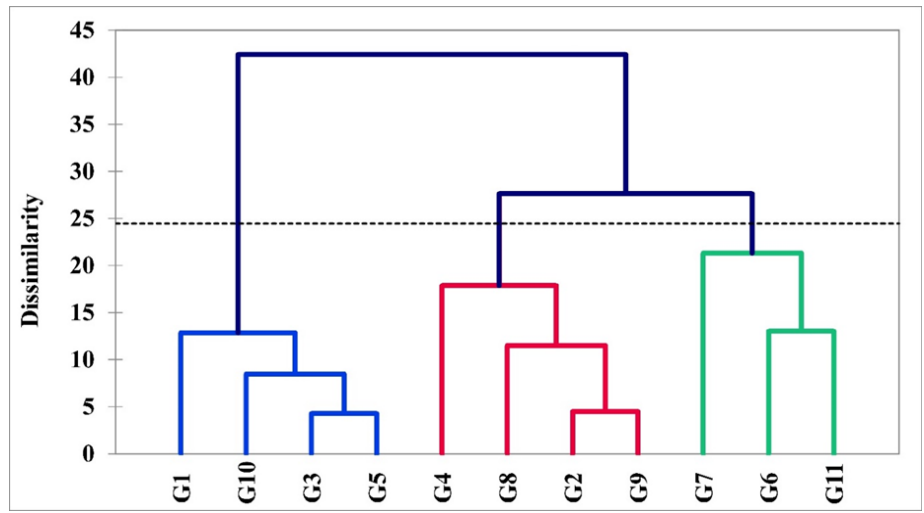

*Figure 6 – The results of Euclidean distance method-based cluster analysis of the studied pomegranate genotypes based on the characteristics of hardwood cuttings.*

## Conclusion

The results obtained from this study showed that the hardwood cuttings of the pomegranate genotypes can be divided into three groups in relation to rooting capacity. Genotypes G6, G7, and G11 had a high rooting percentage (83–95%), genotypes G2, G4, G8, and G9 had a moderate rooting percentage (63–71%), while genotypes G1, G3, G5, and G10 demonstrated a low rooting percentage (28–40%). Based on the results of the irrigation frequencies, a 10-day irrigation frequency reduced the rooting percentage in G1, G2, and G6. Whereas, the lowest rooting percentage was detected at 1-day frequency in G3 and G9. The rooting percentage was not significantly different in G5, G7, G8, G10, and G11 under the effect of the four irrigation frequencies. Besides, irrigation frequencies had a marked effect on root and shoot growth in some but not all of the genotypes.